\documentstyle [12pt,epsfig]{article}
\title{Can a Measurement of the $B_{s} - {\overline{B}}_{s} \,$
\\Mass Difference Establish the CKM Paradigm?}
\author{R. D. Peccei and K. Wang
\\ Dept. of Physics, University of California, Los Angeles
\\ Los Angeles, California 90024}
\date{}
\begin{document}
\def\gsim{\mathrel{\raise.3ex\hbox{$>$\kern-.75em\lower1ex\hbox{$\sim$}}}}
\def\lsim{\mathrel{\raise.3ex\hbox{$<$\kern-.75em\lower1ex\hbox{$\sim$}}}}

\maketitle

\begin{abstract}
We present a reanalysis of the allowed region in the $\rho - \eta \,$ plane of
the CKM matrix, which follows from our present knowledge of the theoretical and
experimental parameters associated with quark mixing and CP violation. Besides
providing updated expectations for the angles of the unitarity triangle, this
reanalysis predicts a range of allowed values for the $B_{s} -
{\overline{B}}_{s} \,$ mass difference $\Delta m_{s}$. We argue that while
values of $\Delta m_{s} \lsim 10 (ps)^{-1}$ could be consistent with a non-CKM
origin for CP violation in the neutral Kaon system, larger values for $\Delta
m_{s}$ would provide strong support for the CKM paradigm.
\end{abstract}

\vspace{4.5cm}
UCLA/95/TEP/2

\pagebreak

It has been realized for a long time \cite{KLS} that a measurement of the
$B_{s} - {\overline{B}}_{s} \,$ mass difference $\Delta m_{s} \,$ (or
equivalently, of the $B_{s} - {\overline{B}}_{s} \,$ mixing parameter
$x_{s}=\Delta m_{s} \tau_{Bs} \,$) would add important information to our
knowledge of the CKM mixing matrix. The equivalent parameters for the $B_{d} -
{\overline{B}}_{d} \,$ complex are now relatively well established, through
measurements of the time integrated mixing probablity at the $\Upsilon(4S)$ -
sensitive to $x_{d}$ - and by direct analysis at LEP of the time dependence of
this mixing - sensitive to $\Delta m_{d}$. A recent summary analysis of all
these measurements by Forty \cite{Forty} - using the world average value for
the $B_{d}$ lifetime, $\tau_{Bd}=1.61 \pm 0.09 (ps)$ \cite{Roudeau}- yields the
values:
\[ \Delta m_{d} = 0.496 \pm 0.032 \, (ps)^{-1}, \, x_{d} = 0.78 \pm 0.05 \, .
\]

Because the mass differences $\Delta m_{d}$ and $\Delta m_{s}$ arise
theoretically from identical box graphs, save for the interchange of d
$\Longleftrightarrow$ s quarks, they are simply interrelated \cite{KLS}.
Indeed, apart from SU(3)-breaking factors associated with evaluating $\Delta
B_{d}=2$ and $\Delta B_{s}=2$ quark operators between $B_{d}$,
${\overline{B}}_{d}$ and  $B_{s}$, ${\overline{B}}_{s}$ states, respectively,
the ratio of these mass differences measures simply the ratio of two CKM matrix
elements:
\begin{equation}
\frac{\Delta m_{s}}{\Delta m_{d}} = \left\{ \frac{M_{Bs}B_{s}f_{Bs}^{2}}
{M_{Bd}B_{d}f_{Bd}^{2}} \right\} \frac{|V_{ts}|^{2}}{|V_{td}|^{2}} \equiv
\xi_{s}^{2} \frac{|V_{ts}|^{2}}{|V_{td}|^{2}} \, .
\end{equation}
The quantity in the curly bracket, which we have denoted by $\xi_{s}^2$ , is
expected to be of order unity. It has been calculated theoretically recently by
various groups using lattice QCD \cite{JS} and from QCD sum rules
\cite{Narison}. In what follows, we shall employ the value that Forty
\cite{Forty} uses:
\[ \xi_{s}^{2} = 1.3 \pm 0.2 \, . \]
This is consistent with the recent theoretical results quoted and also is in
line with the value adopted by Ali and London for this quantity in their recent
analysis \cite{AL}. Using the standard Wolfenstein \cite{Wolf} expansion of the
CKM matrix, one has, to a good approximation, $|V_{ts}| \simeq |V_{cb}|$. Since
this latter matrix element is now reasonably well determined \cite{Stone},
$|V_{cb}|=0.0378 \pm 0.0026$, one sees that a measurement of $\Delta m_{s}$
provides direct information on $|V_{td}|$. Alternatively, without appealing to
a direct value for $|V_{cb}|$, but instead using the Wolfenstein
parametrization of the CKM matrix, a measurement of $\Delta m_{s}$ along with a
knowledge of $\Delta m_{d}$ fixes an allowed region in the $\rho - \eta$ plane:
\begin{equation}
\Delta m_{s} = \Delta m_{d} \frac{\xi_{s}^{2}}{\lambda^{2}} \left[
\frac{1}{(1-\rho)^{2}+\eta^{2}} \right] \, ,
\end{equation}
where $\lambda=0.221 \pm 0.002$ \cite{LR} is the sine of the Cabibbo angle.

There have been recent interesting attempts at LEP to obtain some direct
information on $\Delta m_{s}$ by searching for two frequency components in the
proper-time distribution of tagged $B^{0}$ decays, with results from OPAL
\cite{OPAL} and ALEPH \cite{ALEPH} being presented at the Glasgow conference.
The ALEPH result provides a particularly strong bound on $\Delta m_{s}$ :
\[ \Delta m_{s} > 6 \, (ps)^{-1} \qquad (95\%  \; \rm{C.L.}) \, ,\]
from which, by fluctuating up $1\sigma$ the values of $\Delta m_{s}$ and
$\xi_{s}^{2}$ given, one can infer an upper bound constraint in the  $\rho -
\eta$ plane:
\[\sqrt{(1-\rho)^{2}+\eta^{2}} < 1.61 \, .\]
This bound turns out to be rather close to, or even to somewhat restrict, the
"allowed" region in the $\rho - \eta$ plane determined by various recent
analysis of constraints on the CKM matrix \cite{AL} \cite{Stone} \cite{Rosner}.
However, since the allowed region encompasses values for the above square-root
which are as low as 0.6 - 0.8, it appears that, at present, a very large range
for $\Delta m_{s}$ (and thus also for $x_{s}$ ) is permitted. Nevertheless, as
we will demonstrate below, trying to obtain better bounds on $\Delta m_{s}$
(and certainly measuring this parameter) can provide important insights for the
CKM paradigm.

For these purposes, it is useful to present here a reanalysis of the
constraints on the CKM matrix elements. Basically, three measurements fix the
allowed region in the  $\rho - \eta$ plane: those of $\epsilon$, the CP
violating parameter inferred from neutral Kaon decays; the value of $\Delta
m_{d}$ (or $x_{d}$ ), characterizing $B_{d} - {\overline{B}}_{d}$ mixing; and
the ratio of $|V_{ub}|/|V_{cb}|$, obtained from studing semileptonic B decays
near the end-point region of the electron spectrum. To complete the analysis,
however, further experimental and theoretical information is needed. To
translate the experimental value of $\epsilon$ into a constraint on $(\rho, \,
\eta)$, one needs to know $|V_{cb}|$ and the top quark mass $m_{t}$ , as well
as have a theoretical estimate of the relevant $K_{0} - {\overline{K}}_{0}$
matrix element(which is quantified in terms of the parameter $B_{k}$) . For
$B_{d} - {\overline{B}}_{d}$ mixing, besides needing values for $|V_{cb}|$ and
$m_{t}$, the corresponding $B_{d}
- {\overline{B}}_{d}$ matrix element needed requires a knowledge of $f_{Bd}^{2}
B_{d}$ . Finally, $|V_{ub}|/|V_{cb}|$ can be extracted directly from the data.
However doing so necessitates some model input, and the uncertainties in the
models significantly expand the experimental error.

As the formulas and procedures for relating $\epsilon$ , $\Delta m_{d}$ and
$|V_{ub}|/|V_{cb}|$ to allowed regions in the $\rho - \eta$ plane are fairly
standard \cite{RDP}, we shall not repeat them here. Rather, we give in Table 1
a summary of the values we have used in our analysis. We include in the table
two values of $|V_{ub}|/|V_{cb}|$ , one where uncertainties due to model
dependences are taken into account and the other where this ratio is extracted
using a particular partonic model which we favor - the ACM model \cite{ACM}.
Fig. 1 displays our results, with the cross-hatched area giving the region in
the $\rho - \eta$ plane defined by the three intersecting $1\sigma$ bands,
using the value of $|V_{ub}|/|V_{cb}|$ having the larger uncertainty due to
model dependence. If instead, one uses the ACM model value for
$|V_{ub}|/|V_{cb}|$ , the resulting overlap region - shown as the dashed swath
- is much more restricted already. Expanding the errors slightly and following
some rather standard procedures \c

ite{FJ}, one arrives at the combined allowed 1$\sigma$ region in the $\rho -
\eta$ plane shown in Fig. 2 .

Having determined the allowed region for $(\rho, \eta)$, it is straightforwand
to deduce the parameter range allowed for various quantities of experimental
interest. Principal among these are the $\alpha,\,\beta,\,\gamma$ angles of the
unitary triangle, whose values determine the size of the CP violating
asymmmetries in neutral B decays to CP self-conjugate states \cite{NQ}. Fig. 3
displays the combined allowed range for $sin(2\beta)$ versus $sin(2\alpha)$
which follow from our analysis. Similarly, one can deduce from Fig. 2 and Eq.
(2) an allowed range for $\Delta m_{s}$ \nolinebreak \footnote{To obtain an
estimate of this range, we fluctuate the value of the quantity $(\Delta
m_{d}\xi_{s}^{2}/\lambda^{2})$ from Eq. 2 by $\pm 1 \sigma$ , then combine with
the allowed values of $(\rho, \, \eta)$ from Fig. 2. with errors added in
quadrature.}. This range, as alluded to earlier, is quite large:
\[ 6.2 \, (ps)^{-1} \, < \, \Delta m_{s} \, < \, 22.7 \,  (ps)^{-1} \, . \]
This does not change much even if one restricts oneself only to the region
allowed by the ACM model:
\[ 6.6 \, (ps)^{-1} \, < \, (\Delta m_{s})_{ACM} \, < \, 22.7 \,  (ps)^{-1} \,
. \]

The situation is radically different, however, if one presumes that the
$\epsilon$ - parameter typifying CP violation in the Kaon system - has a non
CKM origin, as for example it does in the superweak model \cite{Wolf 64}. In
this case, one would still have a quark mixing matrix, but this matrix rather
than being unitary, would be orthogonal. In the standard Wolfenstein
parameterization \cite{Wolf} that we are using, this corresponds to having
$\eta \equiv 0$ . Although here $\epsilon$ provides (by assumption) no
constraint, both $B_{d} - {\overline{B}}_{d}$ mixing and the ratio
$|V_{ub}|/|V_{cb}|$ determine a corresponding allowed region for the remaining
free parameter $\rho$ . For the choice of parameters given in Table 1, the
allowed values for $\rho$ for the most probable solution - corresponding to the
overlapping segment on the real axis in Fig. 1 - is given by
\[ \rho = -0.33 \pm 0.08 \qquad (\eta = 0) \, . \]
In this circumstance, the predicted range for  $(\Delta m_{s})$ is considerably
narrower:
\[ 6.0 \, (ps)^{-1} \, < \, \Delta m_{s} \, < \, 9.2 \,  (ps)^{-1} \qquad (\eta
= 0) \, . \]
Indeed, since the allowed value for $\Delta m_{s}$ in this case is almost
entirely dominated by the allowed range for $|V_{ub}|/|V_{cb}|$ , one can write
the approximate equation
\begin{equation}
\Delta m_{s} \, \simeq  \,\Delta m_{d} \frac{\xi_{s}^{2}}{\lambda^{2}}
\frac{1}{(1 + \frac{|V_{ub}|}{|V_{cb}|})^{2}} \qquad (\eta = 0) \, .
\end{equation}
If one were to further restrict oneself to the region allowed by the ACM model,
the above equation would narrow down the allowed range for $\Delta m_{s}$ to
\[ 5.9 \, (ps)^{-1} \, < \, (\Delta m_{s})_{ACM} \, < \, 8.3 \,  (ps)^{-1}
\qquad (\eta = 0) \, . \]

One sees from this discussion, that if a unitary CKM matrix is the correct
paradigm for CP violation, improved bounds on  $\Delta m_{s}$ are unlikely to
make a significant impact on the range of values allowed for the CKM matrix.
However, conversely, a significant improvement of the ALEPH bound to  $\Delta
m_{s} \gsim 10 \, (ps)^{-1}$ , which may be possible at the SLC \cite{MB},
could serve to establish the CKM paradigm by excluding the possibility that
$\eta$ vanishes. In fact, modest improvements in our understanding of the
theoretical uncertainties in both $|V_{ub}|/|V_{cb}|$ and $\xi_{s}^{2}$ could
bring the necessary lower bound on $\Delta m_{s}$ for these purposes down to
about 8.5 $(ps)^{-1}$ .

\bf Acknowledgements: \rm We would like to thank R. Cousins for some helpful
discussions. This work is supported in part by the Department of Energy under
Grant No. FG03-91ER40662.
\newpage

\pagebreak
\begin{table}[h]
\caption{Parameters used for the $\rho \, - \, \eta$ plane analysis}
\vspace{8 mm}
\begin{tabular}{|l|l|}
\hline
$|\epsilon| \, = \, (2.26 \pm 0.02) \times 10^{-3}$ \cite{PDG} &
$m_{t} \, = \, 174 \pm 10  { }_{- 12}^{+ 13} $ GeV  \cite{CDF} \\ \hline
$B_{k} \, = \, 0.825 \pm 0.035 $ \cite{Sharpe} &
$\sqrt{B_d} f_{Bd} \, = \, 180 \pm 30 $ MeV \cite{Guess} \\ \hline
$\Delta m_{d} \, = \, (0.496 \pm 0.032)  \mbox{ ps}^{-1}$ \cite{Forty} &
$|V_{cb}| \, = \, 0.0378 \pm 0.0026 $ \cite{Stone} \\ \hline
$|V_{ub}/V_{cb}| \, = \, 0.08 \pm 0.02 $ \cite{Stone} &
$|V_{ub}/V_{cb}| \, = \, 0.082 \pm 0.006  \mbox{  (ACM)}$ \cite{Stone} \\
\hline
\end{tabular}
\end{table}

\vspace{4cm}
\begin{figure}[h]
{}~\epsfig{file=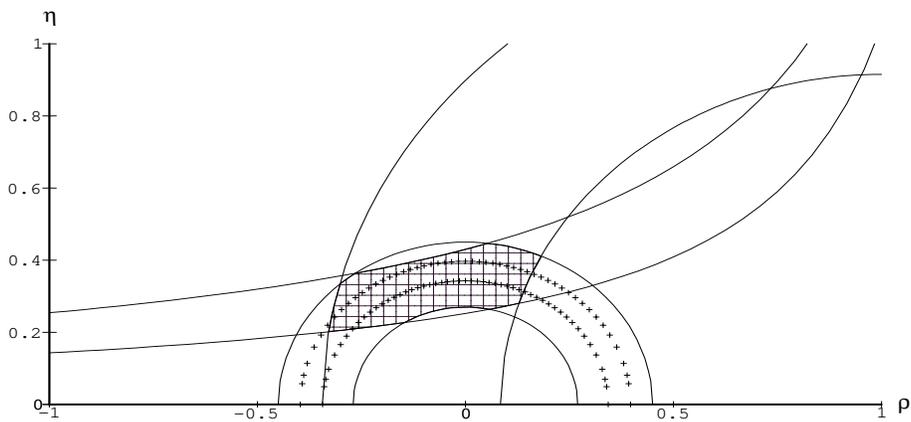,width=13.5cm,height=7.5cm}
\caption{Constraints on the $(\rho, \, \eta)$ plot}
\end{figure}
\newpage

\pagebreak
\begin{figure}[h]
{}~\epsfig{file=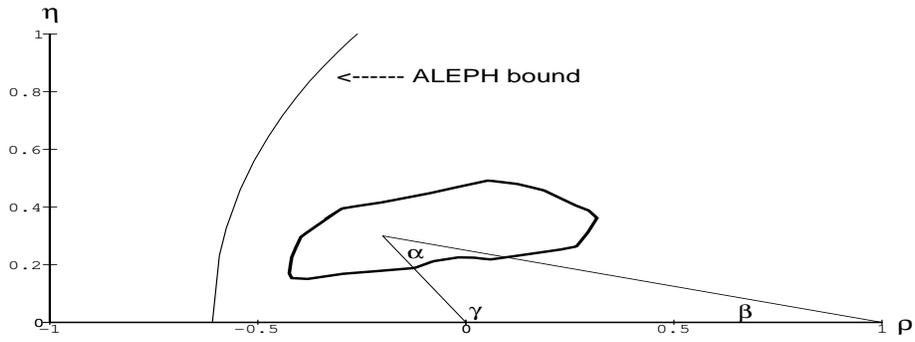,width=13.5cm,height=6cm}
\caption{Allowed region in the $\rho \, - \, \eta$ plane}
\end{figure}

\vspace{3cm}
\begin{figure}[h]
{}~\epsfig{file=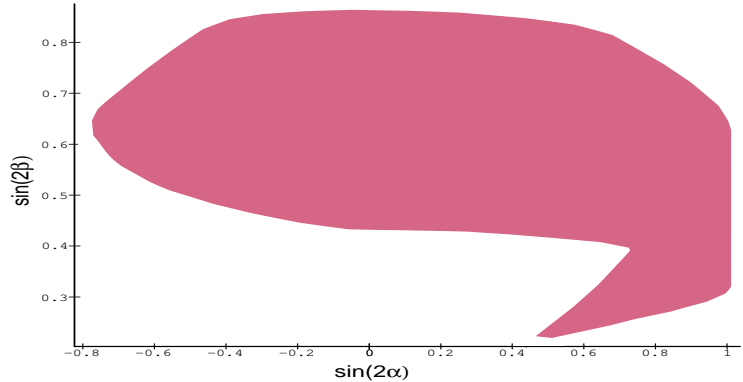,width=13.5cm,height=6cm}
\caption{Allowed region of sin$(2\beta)$ versus sin$(2\alpha)$}
\end{figure}
\newpage

\pagebreak


\begin{thebibliography}{Roudeau}
\bibitem{Guess} This is our guesstimate of the value of this parameter coming
from lattice QCD - for a compilation of the most recent results and a
discussion, see \cite{JS}.
\end{thebibliography}
\end{document}